\documentclass[12pt]{article}
\usepackage[dvips]{graphicx}
\usepackage{textcomp}
\usepackage{amssymb}

\newcommand{\mc}{\multicolumn}

\begin{document}

\begin{titlepage}
\vskip0.5cm
\begin{flushright}
%Berlin\\
\end{flushright}

\vskip0.5cm

\begin{center}
{\Large\bf  The specific heat, the energy density and the thermodynamic
Casimir force in  the neighbourhood of the $\lambda$-transition}
\end{center}

\centerline{
\large Martin Hasenbusch
}
\vskip 0.3cm
\centerline{\sl  Institut f\"ur Physik, Humboldt-Universit\"at zu Berlin}
\centerline{\sl Newtonstr. 15, 12489 Berlin, Germany}
\centerline{\sl
e--mail: \hskip 1cm
 Martin.Hasenbusch@physik.hu-berlin.de}
 \vskip 0.4cm
\begin{abstract}
We discuss the relation of the specific heat, the energy density and the 
thermodynamic Casimir effect  in the case of
thin films in the three dimensional XY universality class.
The finite size scaling function $\theta(x)$ of the thermodynamic Casimir force
can be expressed in terms of the scaling functions $h'(x)$ and $h(x)$ of the 
excess energy density and the excess free energy density.  A priori these 
quantities depend on the reduced temperature $t$ and the thickness $L_0$
of the film. However finite size scaling theory predicts that the scaling 
functions depend only on the combination  $x=t [L_0/\xi_0]^{1/\nu}$, where 
$\nu$ is the critical exponent and $\xi_0$ the amplitude of the correlation 
length.
We exploit this fact to compute $\theta$ from Monte Carlo data for the
excess energy density of the improved two-component $\phi^4$ model 
on the simple cubic lattice with free boundary conditions in the short 
direction.
We repeat this exercise  using experimental data for the 
excess specific  heat of $^4$He films. The finite size scaling 
behaviour of the excess specific heat is governed by $h''(x)$, which 
is proportional to the scaling function $f_2$ discussed in the 
literature.  We compare our results with previous work, where the 
Casimir force has been computed by taking the derivative of the 
excess free energy with respect to the thickness of the film.
As a preparative study we have also computed the scaling functions $h'(x)$ and 
$h(x)$ for finite $L^3$ systems with periodic boundary conditions in 
all directions, where $L$ is the linear extension of the system.
\end{abstract}
\end{titlepage}

\section{Introduction}
In 1978  Fisher and de Gennes \cite{FiGe78} realized that  when thermal
fluctuations are restricted by a container a force acts on the walls
of the container. Since this effect is rather similar to the Casimir effect, 
where the restriction of quantum fluctuations induces a force, it is called 
``thermodynamic'' Casimir effect. Since thermal fluctuations only extend to 
large scales in the neighbourhood of a continuous phase transitions it is 
also called ``critical'' Casimir effect. Recently this effect has attracted 
much attention, since it could be verified for various experimental systems
and quantitative predictions could be obtained from Monte Carlo 
simulations of spin models \cite{Ga09}.

In the thermodynamic limit of the three dimensional system, the correlation 
length, which measures the spatial extension of fluctuations, 
diverges following the power law
\begin{equation}
\label{xipower}
\xi \simeq \xi_{0,\pm} |t|^{-\nu}
 \end{equation}
where $t = (T-T_c)/T_c$ is the reduced temperature and $T_c$ the critical
temperature. $\xi_{0,+} $ and $\xi_{0,-}$ are the amplitudes of the 
correlation length in the high and low temperature phase, respectively.
While $\xi_{0,+} $ and $\xi_{0,-}$ depend
on the microscopic details of the system, the critical exponent $\nu$ and
the ratio $\xi_{0,+}/\xi_{0,-}$ are universal. At the critical point
also other quantities like the  
specific heat show a singular behaviour: 
\begin{equation}
 C \simeq A_{\pm} |t|^{-\alpha} + B \;\;.
\end{equation}
In the case of the XY universality class that we consider here, 
the exponent $\alpha=-0.151(3)$ \cite{recentXY} of the specific heat
is negative. Therefore the analytic background $B$ has to be
taken into account. Note that the critical exponents of the correlation 
length and the specific heat are related by the hyperscaling relation
$\alpha = 2 - d \nu$, where $d$ is the dimension of the system.
For reviews on 
critical phenomena and its modern theory, the Renormalization Group,  
see e.g. \cite{WiKo,Fisher74,Fisher98,PeVi02}.

The singular behaviour at the critical point originates 
from the fact that thermal fluctuations range over all length scales. 
Therefore the behaviour in the neighbourhood of the critical point 
is modified if the system is confined by a container. A priori 
thermodynamic quantities are functions of the reduced temperature 
and the size $L_0$ of the container, assuming a fixed geometry. However 
the theory of finite size scaling 
\footnote{For a review on finite size scaling see \cite{Barber}.}
predicts that the physics of the system is governed by the ratio $L_0/\xi$
as long as $L_0,\xi \gg a$, where $a$ is the microscopic scale of the 
system. In particular if a quantity in the thermodynamic limit behaves as 
$A \simeq a_{0,\pm}  |t|^{-w}$, 
finite size scaling predicts that 
$A(L_0,t)  \simeq  L_0^{w/\nu} \tilde g(L_0/\xi)$,
where $w$ is the critical exponent of $A$ and $\xi$ the correlation length 
of the bulk system. We can rewrite this equation as
\begin{equation}
\label{generalFSS}
 A(L_0,t)  \simeq  L_0^{w/\nu}  g(t [L_0/\xi_0]^{1/\nu})
\end{equation}
by using~(\ref{xipower}), which is the form used in the following. Note
that the function $g$ depends on the details of the container. For
example for a cube it is different from a thin film. It also depends on 
the type of boundary conditions that is imposed by the walls of the 
container on the order parameter of the system.

The predictions of finite size scaling theory have been tested in 
experiments and theoretical studies for various universality classes
and confining geometries; for reviews see \cite{Barber,Privman}.
Here we shall focus on thin films in the three dimensional XY universality 
class, which is shared by the $\lambda$-transition of $^4$He.
Very precise experimental results for critical exponents 
and universal amplitude ratios were obtained for this phase transition
\cite{BaHaLiDu07}.
Also a large number of experiments on thin films of  $^4$He and $^3$He-$^4$He
mixtures were performed to probe finite size scaling \cite{GaKiMoDi08}.
In particular the specific heat $C$ of thin films has been studied.
The excess specific heat  should behave as 
\begin{equation}
C_{bulk}(t)- C(L_0,t) \simeq L_0^{\alpha/\nu} f_2(t [L_0/\xi_0]^{1/\nu}) \;.
\end{equation}
The reason to study the excess specific heat rather than just the 
specific heat $C(L_0,t)$ is to cancel the analytic background $B$.  
Note that the scaling function $f_2(x)$ of the 
excess specific heat is, up to a constant factor, the second derivative 
$h''(x)$ of the scaling function $h(x)$ of the excess free energy per area
\begin{equation}
\label{defineh}
\tilde f_{ex}=\tilde f_{film}(L_0,t)-L_0 \tilde f_{bulk}(t) \simeq k_B T L_0^{-d+1}  h(x)
\end{equation}
where $\tilde f_{film}(L_0,t)$ is the free energy per area of the thin film,
$\tilde f_{bulk}(t)$  the free energy density of the bulk system,
$d=3$ the dimension of the system and $x=t [L_0/\xi_0]^{1/\nu}$. 
Note that in the case of thin films we consider, 
following the literature on the thermodynamic Casimir effect, free energies
per area.  We hope that this does not lead to confusion, since in the 
case of the specific heat, energies per volume are considered.

From a thermodynamic point of view, the Casimir force per unit area is
given by
\begin{equation}
\label{defineF}
 F_{casimir} = - \frac{ \partial \tilde f_{ex} }{ \partial L_0} \;\;
 \end{equation}
where $L_0$ is the thickness of the film. 
Inserting the finite size scaling ansatz~(\ref{defineh}) for the excess 
free energy into (\ref{defineF}) we get
\begin{eqnarray}
\label{mastermind}
 F_{casimir} &\simeq& - k_B T 
 \frac{\partial \left[L_0^{-2}  h(t [L_0/\xi_0]^{1/\nu})]\right]}{ \partial L_0}
\nonumber \\
 &=& - k_B T L_0^{-3} \left[-2 h(t [L_0/\xi_0]^{1/\nu}) + 
 \frac{1}{\nu}  t [L_0/\xi_0]^{1/\nu} h'(t [L_0/\xi_0]^{1/\nu}) \right]
 \nonumber \\
	   &=& k_B T L_0^{-3} \sigma(t [L_0/\xi_0]^{1/\nu})
\end{eqnarray}
where 
\begin{equation}
\label{important}
\sigma(x) = 2 h(x) - \frac{x}{\nu} h'(x) \;. 
\end{equation}
This relation is well known and can be found e.g. in \cite{Krech}. 
We like to emphasis,  that it is at the very heart of finite size
scaling that the behaviour of the Casimir force, which gives the reaction of 
the film with respect to a change of the thickness and the excess specific heat
which gives the reaction of the film with respect to a change of the temperature
are given by the same scaling function $h(x)$ of the excess free energy.

The purpose of the present work is to compute the finite size scaling 
function $\theta(x)$ by using the energy density of thin films obtained 
from Monte Carlo simulations of a lattice model \cite{myheat} and 
by using experimental data for the specific heat of thin films of $^4$He 
near the $\lambda$-transition \cite{KiMeGa99,KiMeGa00}.

As a preliminary study, we simulate $L^3$ systems with periodic 
boundary conditions in all directions.  In order to eliminate leading 
corrections to scaling we have used the improved two component $\phi^4$ 
model on the simple cubic lattice. For a precise definition  of the 
model see section \ref{phi4model}.
Using the results obtained 
for the energy density for a dense grid of temperatures in the 
neighbourhood of the critical point, we investigate the 
scaling behaviour and compute the finite size  scaling functions
$h'(x)$ and $h(x)$. We find that corrections to scaling are small for the 
lattices sizes $L=8$, $16$ and $32$ that we have simulated.

In the case of thin films we analyse data for the energy 
density that were obtained in \cite{myheat} from simulations
of the improved two component $\phi^4$ model on the simple cubic lattice.
In \cite{myheat} these data where used to compute the specific heat.
In order to get a vanishing order parameter as it is observed at
the boundaries of $^4$He films, Dirichlet boundary conditions with
vanishing field were imposed. In singular quantities these lead to  corrections
$\propto L_0^{-1}$ \cite{DiDiEi83}, which can be expressed by an 
effective thickness $L_{0,eff} = L_0 + L_s$.  In \cite{myKTfilm} we find 
$L_s=1.02(7)$ for the model that we consider here. Note that the boundary 
conditions also effect the analytic background of the specific heat 
and the energy density, which also leads to corrections $\propto L_0^{-1}$.
However it turns out that these corrections are not given by the same 
$L_{0,eff}$ as for the singular quantities.
Taking into account these subtleties we arrive at 
accurate result for $h'(x)$, $h(x)$  and $\theta(x)$.  In particular
for $\theta(x)$ in the range $ -15 \lessapprox  x \lessapprox 4 $ 
we find a good match with our previous result \cite{myCasimir}
\footnote{In \cite{myCasimir} we have compared our result for $\theta(x)$ with 
previous ones obtained from simulations of the XY model 
\cite{Hu07,VaGaMaDi08} and  experiments on thin films of
$^4$He \cite{GaCh99,GaScGaCh06}; overall we find a reasonable agreement.},
where we 
computed the Casimir force by taking the derivative of the excess free 
energy with respect to the thickness $L_0$ of the film.

Next we compute $\theta(x)$ by using experimental results for the
excess specific heat obtained from experiments on thin films of 
$^4$He \cite{KiMeGa99,KiMeGa00}. Even though this is a quite simple 
exercise, to our knowledge, it has not been done before.
For $-5 \lessapprox  x \lessapprox 4$ we find a reasonable match 
with our result \cite{myCasimir}. However  in the low temperature phase, 
for $x \lessapprox -5$ we get results that strongly deviate from
 \cite{myCasimir} and can be ruled out by plausibility.
This corroborates the 
observation that in the low temperature phase for $x \lessapprox -5$ 
the excess specific heat does not scale well \cite{GaKiMoDi08}.

This paper is organized as follows: First we define the model and the
observables that we consider. Next we discuss  the finite size
scaling behaviour of the free energy density. In particular, we discuss
corrections to scaling  caused by Dirichlet boundary conditions.
In section \ref{Speriodic} 
we compute the scaling functions $h'(x)$ and $h(x)$ for  $L^3$ systems
with periodic boundary conditions. Then in section \ref{Sfree}
we compute $h'(x)$, $h(x)$ and 
$\theta(x)$ using the data for the energy density of thin films with 
Dirichlet boundary conditions obtained in \cite{myheat}. The result for 
$\theta(x)$ is compared with the one that we  
\cite{myCasimir} obtained directly from the thermodynamic Casimir force.  
Next in section \ref{Sheat}
we compute $\theta(x)$ starting from data for the excess specific heat of 
films of $^4$He in the neighbourhood of the $\lambda$-transition.
Finally we summarize and conclude.

\section{The model and the observables}
\label{phi4model}
We study the two component $\phi^4$ model on the simple cubic lattice.
We  label the sites of the lattice by
$x=(x_0,x_1,x_2)$. The components of $x$ might assume the values
$x_i \in \{1,2,\ldots,L_i\}$. In this work we have performed simulations 
of lattices with $L_0=L_1=L_2$ and periodic boundary conditions in all
three directions. Furthermore we analyse data obtained in \cite{myheat}
for thin films. In this case lattices of the size $L_1=L_2=L$ and 
$L_0 \ll L$ are studied.  In 1 and 2-direction periodic boundary conditions and
free boundary conditions in 0-direction are employed. This means that the sites
with $x_0=1$ and $x_0=L_0$ have only five nearest neighbours.
This type of boundary conditions could be interpreted as Dirichlet
boundary conditions with $0$ as value of the field at $x_0=0$ and $x_0=L_0+1$.
Note that viewed this way, the thickness of the film is $L_0+1$ rather
than $L_0$. This provides a natural explanation of the result $L_s=1.02(7)$
obtained in \cite{myKTfilm}. 
The Hamiltonian of the two component $\phi^4$ model, for a vanishing
external field, is given by
\begin{equation}
\label{hamiltonian}
{\cal H} = - \beta \sum_{<x,y>} \vec{\phi}_x \cdot \vec{\phi}_y
   + \sum_{x} \left[\vec{\phi}_x^2 + \lambda (\vec{\phi}_x^2 -1)^2   \right] 
\end{equation}
where the field variable $\vec{\phi}_x$ is a vector with two real components.
$<x,y>$ denotes a pair of nearest neighbour sites on the lattice.
The partition function is given by
\begin{equation}
Z =  \prod_x  \left[\int d \phi_x^{(1)} \,\int d \phi_x^{(2)} \right] \, \exp(-{\cal H}).
\end{equation}
Note that following the conventions of our previous work, e.g. \cite{ourXY},
we have absorbed the inverse temperature $\beta$ into the Hamiltonian.
\footnote{Therefore, following \cite{Fisher98} we actually should call it
reduced Hamiltonian.}
In the limit $\lambda \rightarrow \infty$ the field variables are fixed to
unit length; i.e. the XY model is recovered. For $\lambda=0$ we get the exactly
solvable Gaussian model.  For $0< \lambda \le \infty$ the model undergoes
a second order phase transition that belongs to the XY universality class.
Numerically, using Monte Carlo simulations and high-temperature series
expansions, it has been shown that there is a value $\lambda^* > 0$, where
leading corrections to scaling vanish.  Numerical estimates of $\lambda^*$
given in the literature are $\lambda^* = 2.10(6)$ \cite{HaTo99},
$\lambda^* = 2.07(5)$  \cite{ourXY} and most recently $\lambda^* = 2.15(5)$
\cite{recentXY}.  The inverse of the critical temperature $\beta_c$ has been
determined accurately for several values of $\lambda$ using finite size
scaling (FSS) \cite{recentXY}. We shall perform our simulations at
$\lambda =2.1$, since for this value of $\lambda$ comprehensive Monte
Carlo studies of the three-dimensional system in the low and the
high temperature phase have been performed
\cite{myKTfilm,recentXY,myAPAM,myamplitude}.
At $\lambda =2.1$ one gets $\beta_c=0.5091503(6)$ \cite{recentXY}.
Since  $\lambda =2.1$  is not exactly equal to $\lambda^*$, there are
still corrections $\propto L^{-\omega}$, although with a small amplitude.
In fact, following \cite{recentXY}, it should be by at least a factor
20 smaller than for the standard XY model.

In \cite{myKTfilm} we find for $\lambda=2.1$ by fitting the data for the
second moment correlation length in the high temperature phase 
$ \xi_{2nd} \simeq 0.26362(8) t^{-0.6717}$, where $t=0.5091503-\beta$.
We shall use this definition of the reduced
temperature also in the following discussion of our numerical results; Hence
$\xi_0=0.26362(8)$.
Note that in the high temperature phase there is little difference between
$\xi_{2nd}$ and the exponential correlation length $\xi_{exp}$ which
is defined by the asymptotic decay of the two-point correlation function.
Following  \cite{ourXY}
$\lim_{t\searrow 0} \frac{\xi_{exp}}{\xi_{2nd}} = 1.000204(3)$
for the thermodynamic limit of the three-dimensional system.
Hence at the
level of precision reached here it does not matter whether $\xi_{0,exp}$ or
$\xi_{0,2nd}$ is used in the scaling variable $x = t [L_0/\xi_0]^{1/\nu}$.

\subsection{The internal energy and the free energy}
\label{defineE}
The reduced free energy density is defined as
\begin{equation}
\label{fdef1}
f = - \frac{1}{L_0 L_1 L_2} \log Z \;.
\end{equation}
I.e. compared with the free energy density $\tilde f$, a factor $k_B T$ is 
skipped.

Note that in eq.~(\ref{hamiltonian})  $\beta$ does not
multiply the second term. Therefore, strictly speaking, $\beta$ is not
the inverse of $k_B T$.
In order to study universal quantities it is not crucial how the transition
line in the $\beta$-$\lambda$ plane is crossed, as long as this path is
not tangent to the transition line.
Therefore, following computational convenience, we vary $\beta$ at fixed 
$\lambda$.
Correspondingly we define the (internal) energy density as the derivative of 
the reduced free energy density with respect to $\beta$.
Furthermore, to be consistent with our previous work \cite{myheat}, 
we multiply by $-1$:
\begin{equation}
\label{Edef1}
E = \frac{1}{L_0 L_1 L_2} \frac{\partial \log Z}{\partial \beta} \;.
 \end{equation}
 It follows
\begin{equation}
\label{Edef}
 E =  \frac{1}{L_0 L_1 L_2}
\left \langle  \sum_{<x,y>} \vec{\phi}_x \cdot \vec{\phi}_y \right \rangle \;,
\end{equation}
which can be easily determined in Monte Carlo simulations.  From
eqs.~(\ref{fdef1},\ref{Edef1}) it follows that the free energy density 
can be computed as
\begin{equation}
\label{integrateF}
 f(\beta) = f(\beta_0) - \int_{\beta_0}^{\beta} 
                       \mbox{d} \tilde \beta   E(\tilde \beta)   \;\;.
\end{equation}

\section{The finite size scaling behaviour of the free energy}
\label{theory}
Let us briefly discuss the scaling behaviour of the reduced excess 
free energy per area.   
Since we study an improved model we ignore corrections
$\propto L_0^{-\omega}$ in the following.  We take into account leading
corrections due to the boundary conditions by replacing the thickness
$L_0$ of the film by $L_{0,eff}=L_{0}+L_s$ at the appropriate places. We 
split the free energies in singular (s) and non-singular (ns) parts:
\begin{eqnarray}
 f_{ex}(t,L_0) &=&  f_{film}(t,L_0) - L_0 f_{bulk}(t) \nonumber  \\
 &=& f_{film,s}(t,L_0) + L_{0,eff,ns} f_{ns}(t) - L_{0} f_{bulk,s}(t) 
 - L_0 f_{ns}(t)  \nonumber  \\
 &=& L_{0,eff}^{-2} h(x)
                          + L_s(t) f_{bulk,s}(t)
                          + L_{sns}(t) f_{ns}(t) 
\end{eqnarray}
where 
\begin{equation}
\label{Funiversal}
h(x)=L_{0,eff}^2 [f_{film,s}(t,L_0) - L_{0,eff} f_{bulk,s}(t)]
\end{equation}
is a universal finite size scaling function and 
$x=t [L_{0,eff}/\xi_0]^{1/\nu}$. Following RG theory the non-singular part
is not affected by finite size effects. However it is not clear a priori 
how Dirichlet boundary conditions affect the non-singular part of the 
free energy. Therefore we allow for $L_{sns}=L_{0,eff,ns} - L_{0} \ne 0$ and
$L_{sns} \ne L_{s}$.
Taking the derivative with respect to $L_0$ we get the
thermodynamic Casimir force per area \cite{Krech}
\begin{equation}
\beta F_{casimir} = - \frac{\partial f_{ex}(t,L_0)}{\partial  L_0}=
 2  L_{0,eff}^{-3} h(x)
  - L_{0,eff}^{-3} \frac{1}{\nu} x h'(x)
  = L_{0,eff}^{-3} \theta(x) 
 \end{equation}
where  
\begin{equation}
\label{master}
\theta(x) = 2 h(x) - \frac{1}{\nu} x h'(x)  \;.
\end{equation}
Note that the boundary terms  $L_s f_{bulk,s}$
and $L_{sns} f_{ns}$  do not contribute to the Casimir force. 

\section{Warmup exercise: finite cubic system with periodic boundary conditions}
\label{Speriodic}
First we have studied a finite lattice with $L=L_0=L_1=L_2$  with periodic 
boundary conditions in all directions. 
This way we avoid corrections caused by the free boundary conditions and  
possible 
difficulties  related with the Kosterlitz-Thouless  transition of thin 
films. In our simulations we determine the energy density $E$ for the 
lattice sizes $L=8$, $16$ and $32$ for a large number of $\beta$-values 
in the neighbourhood of the critical temperature. In particular, 
we have simulated at 159 $\beta$-values in the interval 
$0.35 \le \beta \le 0.58$, 205 $\beta$-values in the interval 
$0.35 \le \beta \le 0.58$ and 161 $\beta$-values in the interval
$0.45 \le \beta \le 0.535$ in the case of $L=8$, $16$ and $32$, respectively.
For most of these simulations we performed $10^6$ measurements. 
For each of these measurements we performed one Metropolis sweep, two 
overrelaxation sweeps and a number of single cluster \cite{wolff} updates. The 
number of single cluster updates is chosen such that the number of 
updates times the average size of a cluster roughly equals $L^3$.
As random number generator we have used the  SIMD-oriented Fast
Mersenne Twister algorithm \cite{twister}.
In total we have used about 3 weeks of CPU-time on a single core of a
Quad-Core Opteron(tm) 2378 CPU (2.4 GHz).  In order to compute the 
excess energy density
\begin{equation}
E_{ex}(L,t) = E(L,t) - E_{bulk}(t) 
\end{equation}
we have used the results for $E_{bulk}(t)$ obtained in section 4.1 
of \cite{myheat}. 

\begin{figure}
\begin{center}
\scalebox{0.62}
{
\includegraphics{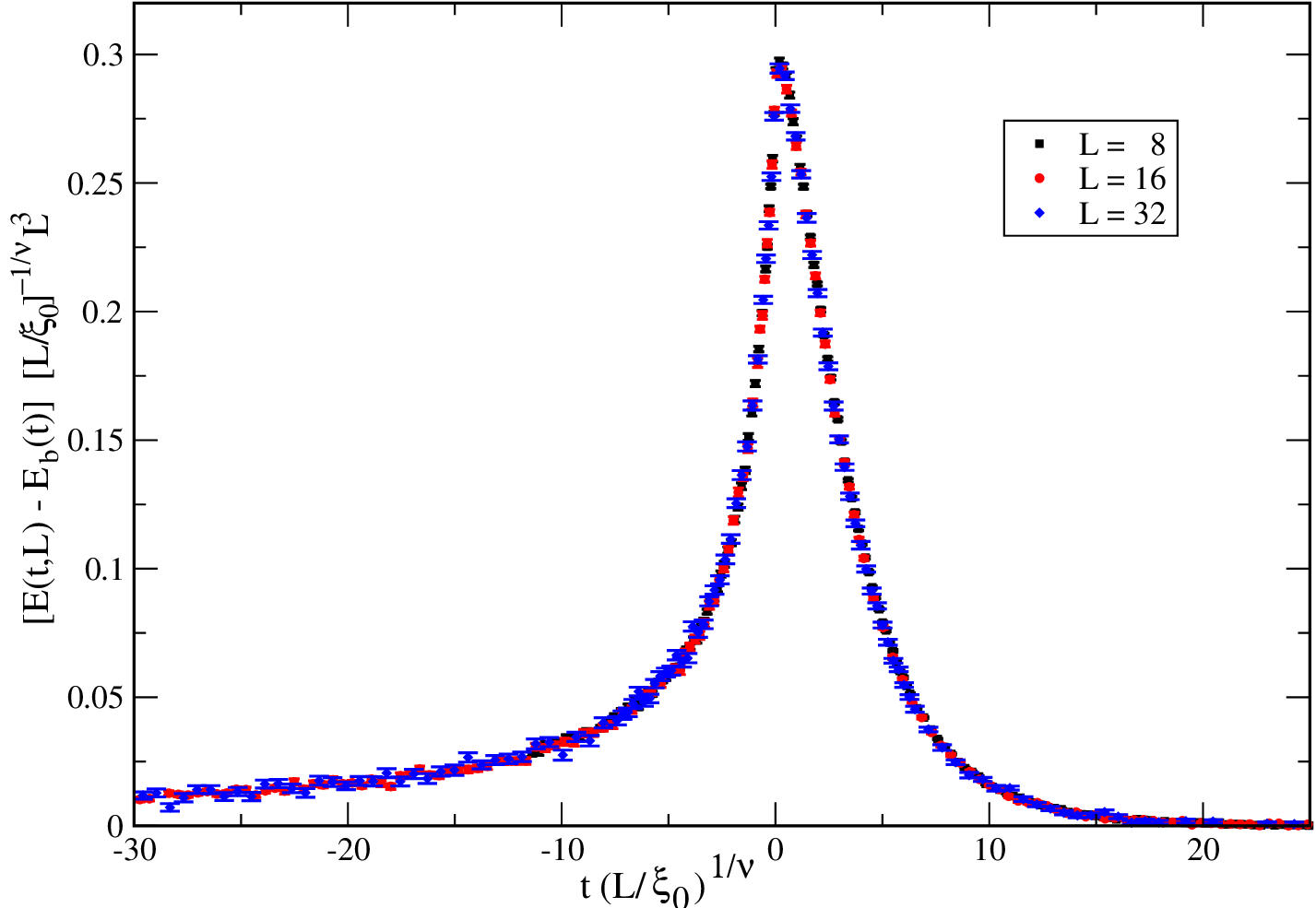}
}
\end{center}
\caption{
\label{Eplot} We consider an $L^3$ system with periodic 
boundary conditions in all directions. We plot 
$[E(t,L) - E_{bulk}(t)] [L/\xi_0]^{-1/\nu}  L^3 $
as a function of $ t (L/\xi_0)^{1/\nu}$ for $L=8$,  $16$ and $32$,
where we use $\nu=0.6717$ and $\xi_0 =0.26362$.
For a discussion see the text.
}
\end{figure}

The reduced excess  free energy density behaves as 
\begin{equation}
\label{hp}
 f(t,L) - f_{bulk}(t) \simeq L^{-3} h(x)
\end{equation}
where $x=t [L/\xi_0]^{1/\nu}$. 
It follows for the excess energy density
\begin{equation}
\label{hpp}
 E(t,L) - E_{bulk}(t) \simeq [L/\xi_0]^{1/\nu}  L^{-3} h'(x) \;.
\end{equation}
Since we study an improved model and periodic boundary conditions 
in all directions, we expect that  analytic corrections are  leading.
In figure \ref{Eplot} we plot 
$[L/\xi_0]^{-1/\nu}  L^3 [E(t,L) - E_{bulk}(t)]$
as a function of $t [L/\xi_0]^{-1/\nu}$. The curves for $L=8$, $16$ and $32$
fall nicely on top of each other, showing that corrections to scaling are 
numerically small. The excess energy is positive for all temperatures. 
At $x \approx - 0.3$ the function assumes a maximum. The value at the 
maximum is $\approx 0.295$. In the high temperature 
phase, for increasing $x$ the function $h'(x)$ rapidly approaches 
zero. In contrast, in the low temperature phase it is only slowly approaching
zero  with decreasing $x$. This behaviour might be explained by the presence 
of a  Goldstone mode in the low temperature phase.

\begin{figure}
\begin{center}
\scalebox{0.62}
{
\includegraphics{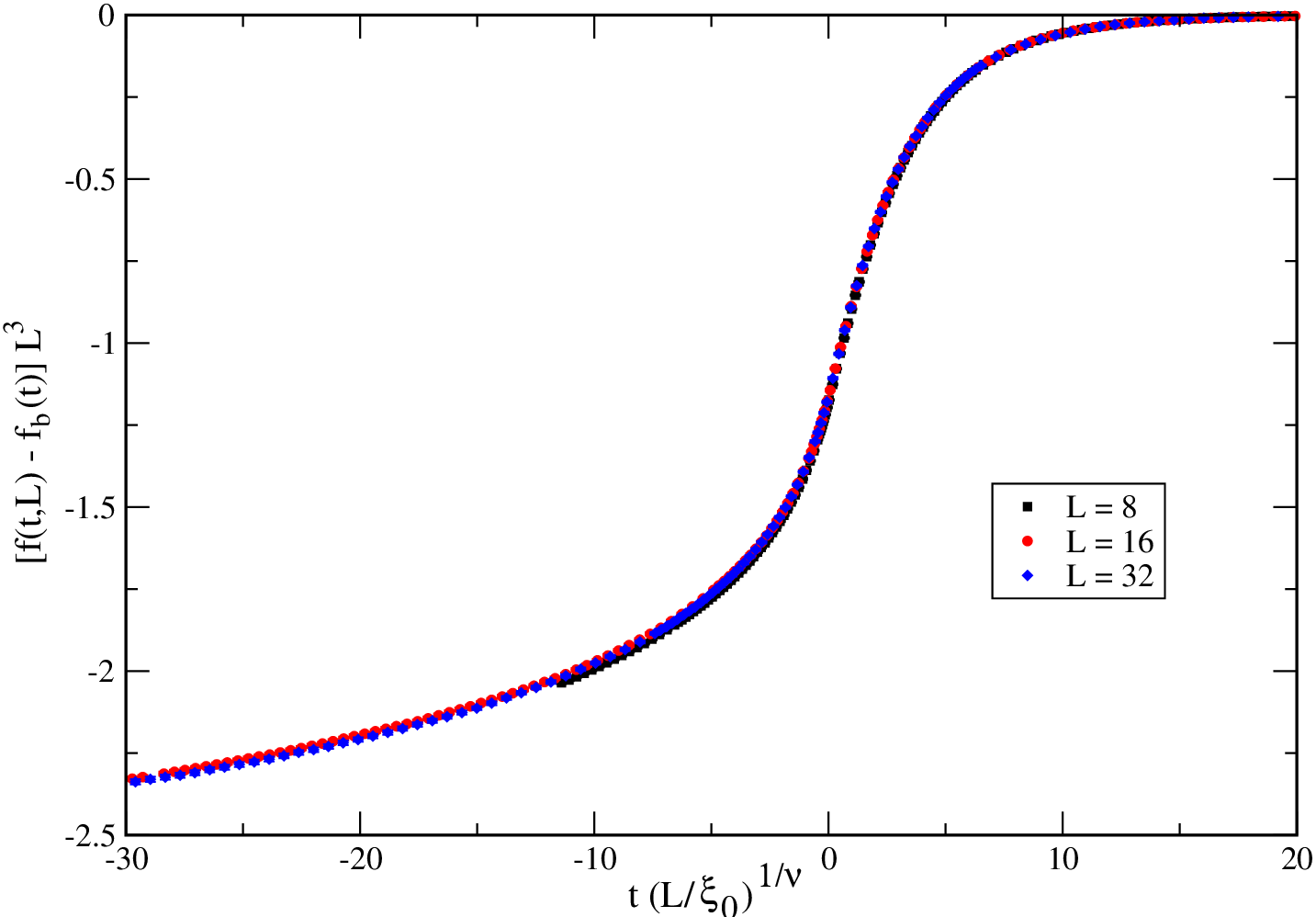}
}
\end{center}
\caption{
\label{Fplot} 
We consider an $L^3$ system with periodic
boundary conditions in all directions. 
We plot $L^3 [f(t,L) - f_{bulk}(t)]$
as a function of $  t (L_0/\xi_0)^{1/\nu}$
for $L_0=8$,  $16$ and $32$, where we use $\nu=0.6717$ and
$\xi_0 =0.26362$.
For a discussion see the text.
}
\end{figure}

Next we have computed the excess free energy using eq.~(\ref{integrateF}).
To this end we integrated 
our data for the excess energy by using the trapezoidal rule. We have 
started the integration at $\beta_0=0.35$, $0.45$ and $0.49$ for 
$L=8$, $16$ and $32$, respectively. At these values of $\beta_0$ the 
deviation of the excess energy from zero is of similar size as the statistical
error.
In figure \ref{Fplot} we plot $L^3 [f(t,L) - f_{bulk}(t)]$
as a function of $t [L/\xi_0]^{-1/\nu}$. As one should expect, the curves 
for $L=8$, $16$ and $32$ fall nicely on top of each other.  Since 
the excess energy is positive for all values of $x$, the excess free
energy is monotonically increasing with increasing $x$.  At the 
critical point the scaling functions assumes the value $h(0) = -1.162(4)$.

In Fig. 6 of \cite{DohmRev} results for an isotropic $L^3$ system with 
periodic boundary conditions for the Ising universality class and the 
limit $n \rightarrow \infty$ of $O(n)$ symmetric models are given.
The Ising result is obtained from a perturbative approach in three 
dimensions fixed, while the $n \rightarrow \infty$ result is exact.
In the Ising case, $h(x)$ shows a minimum in the low temperature phase, 
while for $n \rightarrow \infty$ it is monotonically decreasing as 
the temperature decreases. Hence qualitatively the behaviour for $n=2$, studied
here, is the same as for $n \rightarrow \infty$.

\section{Film geometry with free boundary conditions}
\label{Sfree}
Here we study thin films with free boundary conditions in the short direction.
This geometry is relevant for the comparison with experimental results 
obtained for thin films of $^4$He. 
Most of the Monte Carlo data are 
taken from our previous work \cite{myheat,myCasimir}, where we have simulated
films of the thicknesses $L_0=8$, $16$ and $32$. Therefore we refrain from
giving the details of the simulations and refer the reader to  
\cite{myheat,myCasimir}.

Analogous to the previous section we compute the scaling function 
$h'(x)$ of the 
excess energy density and $h(x)$ of the excess free energy density.
Using these we obtain the scaling function 
$\theta(x)=2 h(x) - \frac{x}{\nu} h'(x)$ of the thermodynamic 
Casimir force.

In \cite{myheat} we have taken great care to get the deviations of the energy 
density from its effectively two dimensional thermodynamic limit
under control. In order to achieve this, quite large ratios $L_1/L_0$ are needed
in the neighbourhood of the peak of the specific heat.
In the case of 
$L_0=8$ we have simulated lattices of a size up to $L_1=L_2=2048$,  and
for $L_0=16$ up to $L_1=L_2=1800$.  For $L_0=32$ we have skipped the 
interval $0.5136 < \beta < 0.516$, since we could not simulate 
sufficiently large values of $L_1=L_2$.

\begin{figure}
\begin{center}
\scalebox{0.62}
{
\includegraphics{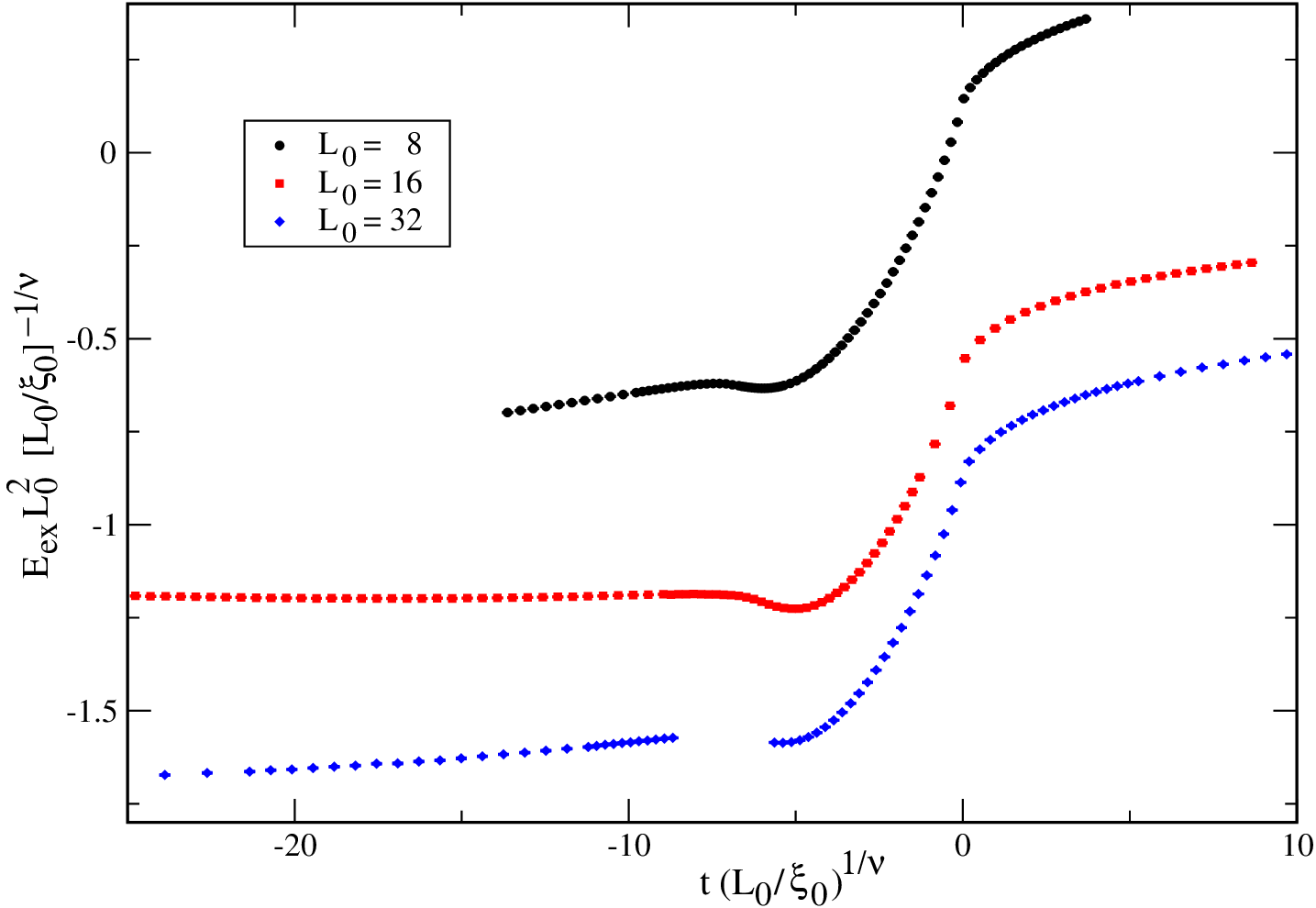}
}
\end{center}
\caption{
\label{naivplot}  We plot $E_{ex} L_0^2 [L_0/\xi_0]^{-1/\nu}$ as a function
of $t [L_0/\xi_0]^{1/\nu}$ for thin films,  where we have used 
$\xi_0 =0.26362$ and $\nu=0.6717$. For a discussion see the text.
}
\end{figure}

In figure \ref{naivplot}, 
similar to the previous section, we have plotted
$E_{ex} L_0^2 [L_0/\xi_0]^{-1/\nu}$  versus $t [L_0/\xi_0]^{1/\nu}$,
where now
\begin{equation}
 E_{ex}(L_0,t) = L_0 [E(L_0,t) - E_{bulk}(t) ]
\end{equation}
is the excess energy per area. 
In contrast to the previous section we find that there is a huge discrepancy 
between the three curves. The dominant effect seems that the curves 
are shifted by a constant with respect to each other. Note that
replacing $L_0$ by $L_{0,eff}=L_0+L_s$ with $L_s=1.02(7)$ \cite{myKTfilm}
does change this situation only little.  In section \ref{theory}  
we have argued  that the analytic background of the energy density might
suffer from a boundary correction that is not given by the effective thickness
$L_{0,eff}=L_0+L_s$ that describes the leading boundary corrections of 
singular quantities. Below we shall study this question in detail 
at the critical point of the bulk system, where we have data for  
thicknesses up to $L_0=64$ available.

\subsection{Finite Size Scaling at the critical point of the bulk system}
In order to get a better understanding of the corrections we have 
studied in detail the behaviour at the critical point of the three
dimensional bulk system. In the context of \cite{myheat} we have simulated
lattices of the thickness $L_0 = 8$, $12$, $16$, $24$, $32$, $48$ and 
$64$ and $L_1=L_2=6 L_0$.  In \cite{myheat} we have checked that this choice 
of $L_1$, $L_2$ is sufficient to approximate well the effectively two 
dimensional thermodynamic limit of the film at the critical point of the 
three dimensional system. 
Our  results for the energy density are summarized in table 
\ref{ENE6}.

\begin{table}
\caption{\sl \label{ENE6} Energy density of thin films of the 
thickness $L_0$ at the inverse critical temperature 
$\beta_c=0.5091503(6)$ of the three dimensional systems. 
In all cases $L_1=L_2=6 L_0$. }
\begin{center}
\begin{tabular}{|r|l|}
\hline
 \mc{1}{|c}{$L_0$}  &   \mc{1}{|c|}{$E$}      \\
\hline
  8   &   0.799566(31) \\ 
 12   &   0.832786(19) \\ 
 16   &   0.850727(13) \\  
 24   &   0.8698028(85) \\ 
 32   &   0.8798552(57) \\
 48   &   0.8903321(37) \\ 
 64   &   0.8957662(29) \\ 
\hline
\end{tabular}
\end{center}
\end{table}

In the case of periodic boundary conditions in all directions, 
the energy density at the critical point behaves as
\begin{equation}
\label{Eperiodic}
 E(L) = E_{ns} + c L^{-d+1/\nu}  \;
\end{equation}
where $d=3$ is the dimension of the system.
Using lattices of the size $L_0=L_1=L_2$  with $L_0$ up to $128$ we find
\cite{myAPAM}
\begin{equation}
\label{periodicEbc}
 E_{ns}= 0.913213(5) + 20 \times  (\beta_c - 0.5091503) 
                     +  5 \times 10^{-7} \times (1/\alpha +1/0.0151) \;.
\end{equation}

In order to take into account corrections due to the free boundary conditions
of the thin film we use the ansatz 
\begin{equation}
\label{Efree}
  L_0 E(L_0) = (L_0 + L_{sns}) E_{ns} + c_f L_{0,eff}^{-d+1+1/\nu}  
\end{equation}
to fit the data given in table \ref{ENE6}. As input we have used 
$\nu=0.6717(1)$ \cite{recentXY}, $E_{ns}$ given in eq.~(\ref{periodicEbc}) 
and $L_s=1.02(7)$ \cite{myKTfilm}.  The parameters of the fit are 
$L_{sns}$ and $c_f$. 
Fitting all data with $L_0 \ge 8$ we get an acceptable $\chi^2/$d.o.f. 
We find $L_{ns} = -1.3529(3) $ when fixing
 $L_{s} = 1.02$ and  $L_{ns} = -1.3523(3)$ fixing $L_{s} = 0.95$.  Hence
$L_{sns}$ is clearly different from $L_{s}$ and it shows little dependence 
on the value taken for  $L_{s}$. We have checked that the error of
$L_{sns}$ due to the uncertainties of $\nu$ and $E_{ns}$  can be ignored.

\subsection{Taking into account boundary corrections}
\begin{figure}
\begin{center}
\scalebox{0.62}
{
\includegraphics{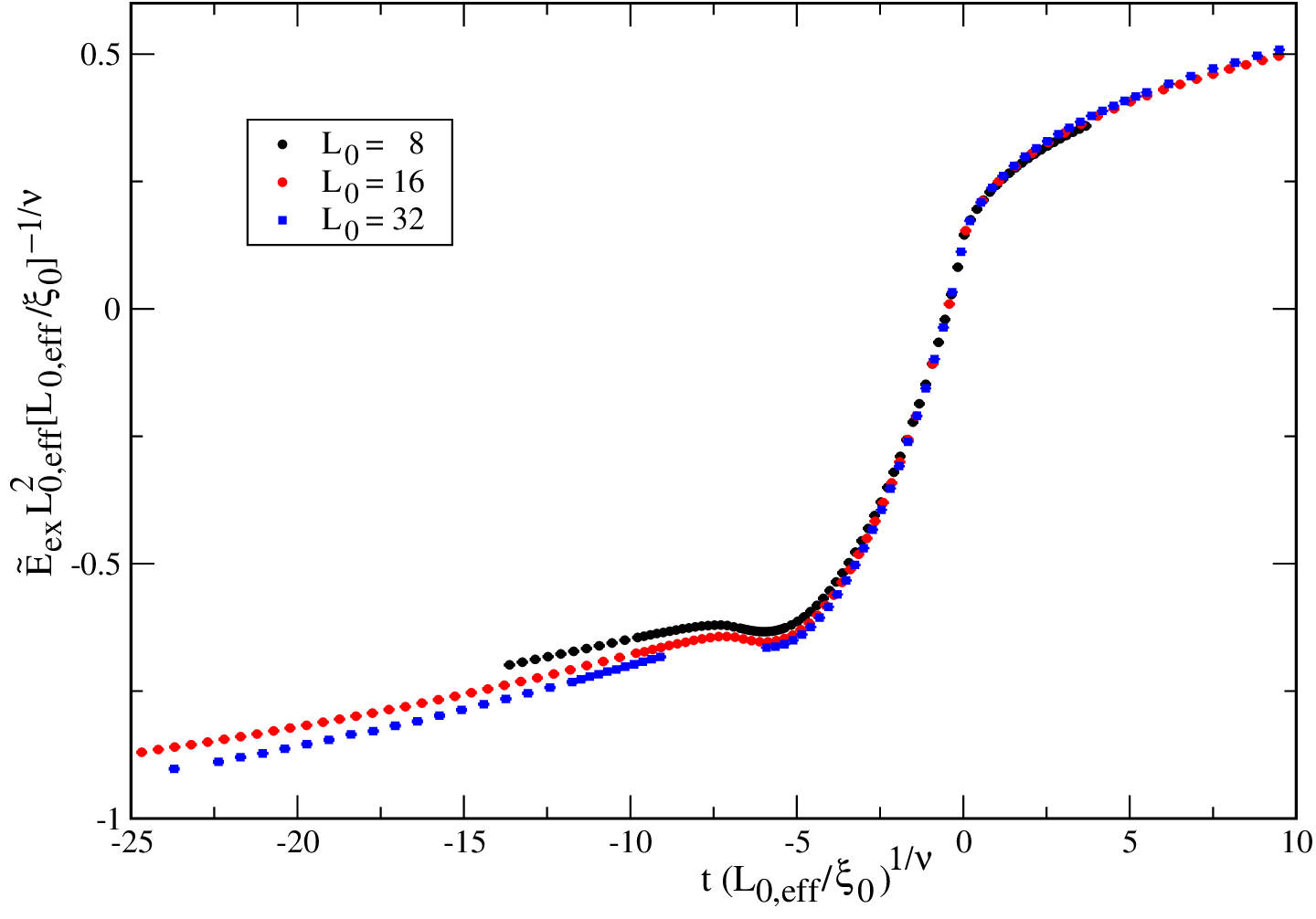}
}
\end{center}
\caption{
\label{scaleplot}  We plot 
$\tilde E_{ex} L_{0,eff}^2 [L_{0,eff}/\xi_0]^{-1/\nu}$
as a function of $t [L_{0,eff}/\xi_0]^{1/\nu}$ for $L_0=8$, $16$ and 
$32$.
For a discussion see the text.
}
\end{figure}

The boundary correction $L_{ns} E_{ns}$ should be an  analytic function of 
the reduced temperature. In a first attempt  we shall approximate it
by its value at the critical point of the three dimensional system found 
above. Hence in figure \ref{scaleplot} we plot 
$\tilde E_{ex} L_{0,eff}^2 [L_{0,eff}/\xi_0]^{-1/\nu} $ where
\begin{equation}
\label{besserEx}
 \tilde E_{ex} = L_0 E(L_0,t) - L_{0,eff} E_{bulk}(t) + (L_s-L_{sns}) E_{ns} 
\end{equation}
as a function of $x=t [L_{0,eff}/\xi_0]^{1/\nu}$.  Now we see a quite 
good matching of the three curves. Only for small $x$ discrepancies are visible.

Next we have computed the finite size scaling function $\theta$ of the 
Casimir force following eq.~(\ref{master}). In the case of 
$L_0=8$ and $L_0=16$ we have used the function $h'(x)$ as given in figure
\ref{scaleplot}. In the case of $L_0=32$ we have taken the missing part 
in the range $-5.9 > x > -9.1$ from the results for $L_0=16$. To this end we 
have matched the values of the function at $x=-5.9$ and $x=-9.1$  resulting
in $h'_{32}(x) = h'_{16}(x) + 0.011 - 0.002 (x+5.9)$ for
$-5.9 > x > -9.1$.
We have computed the function $h(x)$ by numerically integrating $h'(x)$ using 
the trapezoidal rule.  
For sufficiently large $x$ the Casimir force vanishes and therefore 
$h(x) = \frac{x}{2 \nu} h'(x)$.  Hence for large $x$:
\begin{equation}
\label{high}
 h'(x)  =  c x^{2 \nu -1}  \;\;.
\end{equation} 
In  \cite{myCasimir}  we found that the thermodynamic  Casimir force is of
similar size or smaller than the numerical errors that we achieve for
$x \gtrapprox 4$.  We have checked that in this range the scaling 
function $h'(x)$ of the excess energy indeed follows eq.~(\ref{high}).

\begin{figure}
\begin{center}
\scalebox{0.62}
{
\includegraphics{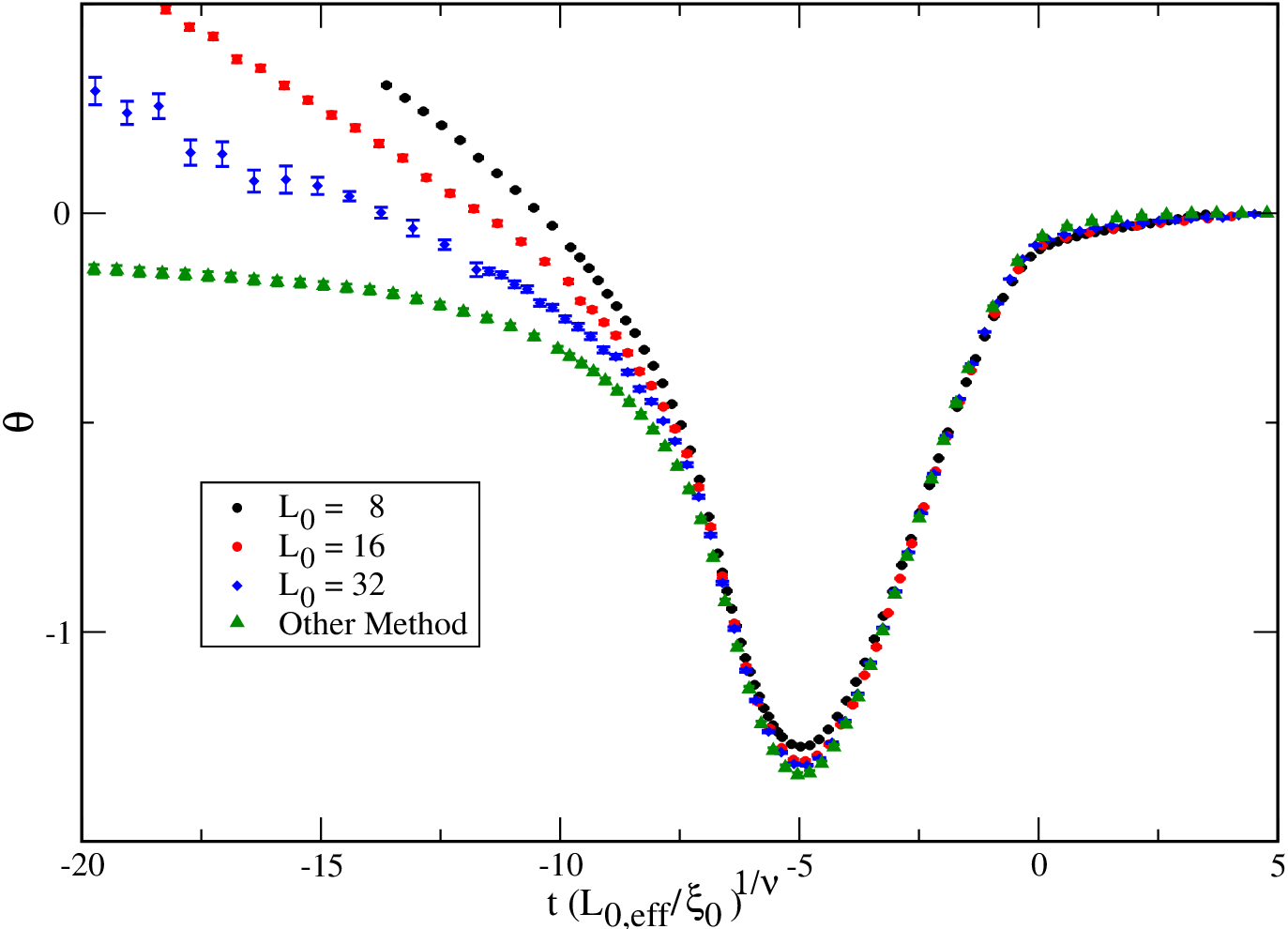}
}
\end{center}
\caption{
\label{casimir1plot}  We plot finite size scaling function $\theta$ of the 
thermodynamic Casimir force. We have computed $\theta$  from the finite 
size scaling function $h'$ of the excess energy per area of the film.
For comparison we give the result of our previous work \cite{myCasimir}
where we have computed $\theta$  directly from the thermodynamic Casimir 
force.  For a discussion see the text.
}
\end{figure}

Hence we have started the numerical integration in the high temperature phase 
at $x_0 \approx 4$ with the starting value $h(x_0)=\frac{x_0}{2 \nu} h'(x_0)$.
In order to check the reliability of our result, 
we have redone the integration using a set of data points, where 
we have skipped every second value of $\beta$. We found an agreement within 
the statistical errors.
Our results for $\theta$ are plotted in figure \ref{casimir1plot}.
In the range $-7 < x < 5$ the curves obtained from the different values of
$L_0$ match quite well. There is also a good match with $\theta$ obtained 
in \cite{myCasimir}. In particular the value and the position of the minimum 
of $\theta$ are completely consistent. However for $x < -7$ the difference 
between the curves becomes clearly visible and increases with decreasing $x$.
For small $x$, even for $L_0=32$ there is a huge discrepancy with  the 
result of \cite{myCasimir}.

\begin{figure}
\begin{center}
\scalebox{0.62}
{
\includegraphics{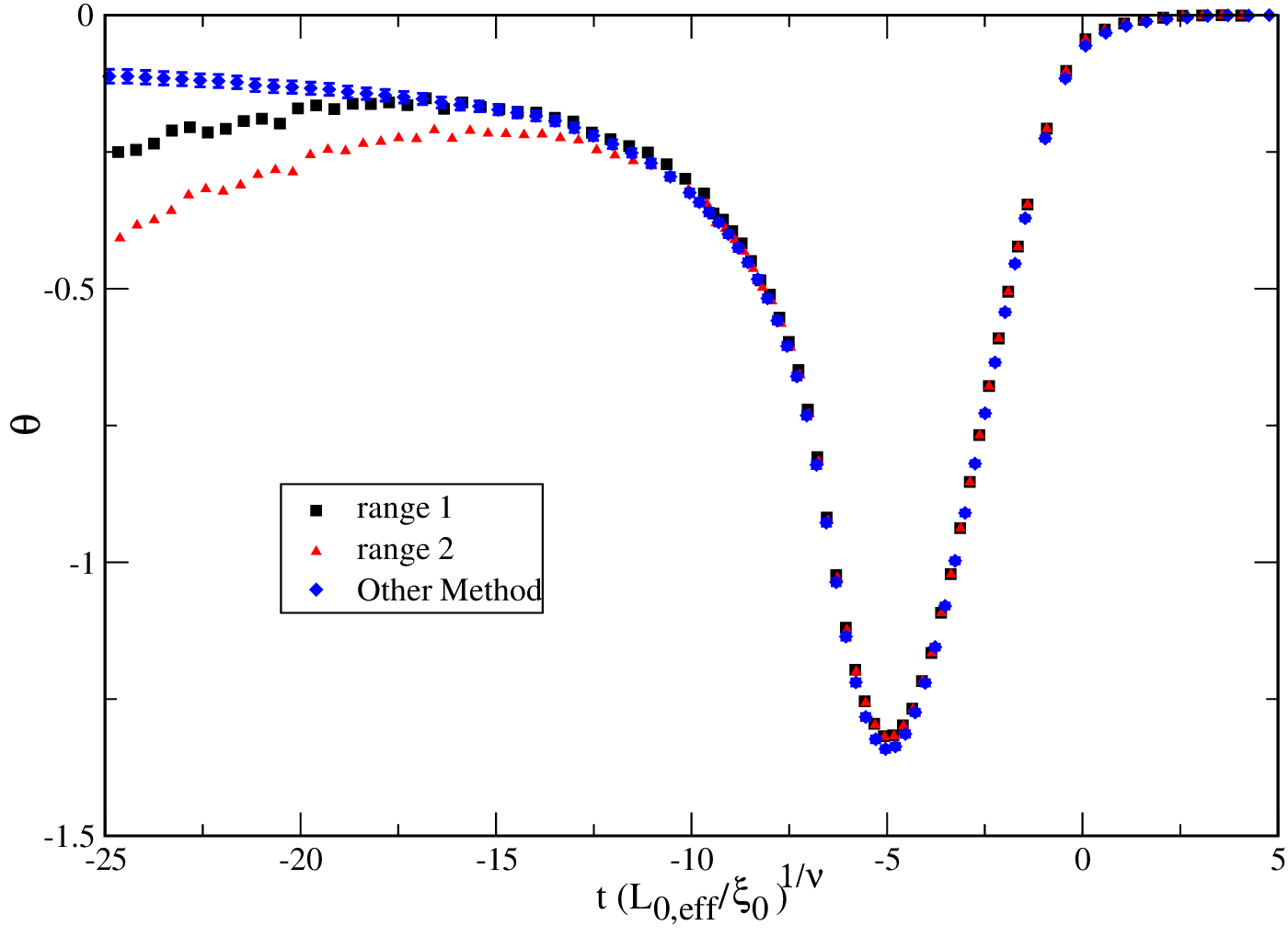}
}
\end{center}
\caption{
\label{casimirmatchplot}  Similar to the previous figure. Here we have
taken into account analytic correction computing the scaling function 
$h'$. For a discussion see the text.}
\end{figure}

Since these discrepancies appear for large values of $|x|$ it is  likely
that they are mainly caused by analytic corrections. To check this 
explicitly, we allowed for two different types of corrections:
\begin{equation}
 x = t (1- c t) [L_{0,eff} /\xi_0]^{1/\nu}
\end{equation}
and for a temperature dependence of the boundary 
correction of the analytic part of the energy
\begin{equation}
\label{besserEx}
 \tilde E_{ex} = L_0 E(L_0,t) - L_{0,eff} E_{bulk}(t) + (L_s-L_{sns}) E_{ns}
 -c_b t  \;\;.
\end{equation}
We find that the curves for $h'(x)$ obtained from $L_0=16$ and $L_0=32$ 
can be nicely matched by adjusting the two parameters $c$ and $c_b$.
Matching in the interval $-18  < x <  3$  we find 
$c \approx -1.1$ and $c_b\approx -3.03$  and for the interval $-25  < x <  5$
$c \approx  -0.75$ and $c_b\approx -2.97$. Using the corresponding results
for $h'(x)$ we have computed the  finite size scaling function $\theta(x)$ that
is plotted in figure \ref{casimirmatchplot}.
Now we see that the rage of the matching with 
our previous result \cite{myCasimir} is extended to 
$-15 \gtrapprox x \gtrapprox  4$ 
in the case of the matching range $-25  < x <  4$. 
For still smaller values of $x$ discrepancies  rapidly increase.
Likely higher order analytic corrections are the main reason for this 
behaviour. However also other types of corrections
like $t^{\nu \omega'}$ with $\omega' \approx 1.8$  \cite{NR}
should be taken into account.
Therefore we abstain from fitting with $t^2$ corrections.

\section{The specific heat of thin films of $^4$He and the thermodynamic 
Casimir force}
\label{Sheat}
In a number of experiments the excess specific heat of thin films of $^4$He 
and $^3$He-$^4$He mixtures
has been measured in the neighbourhood of the $\lambda$-transition
\cite{GaKiMoDi08}. In these works the scaling function $f_2$ which is defined
by
\begin{equation}
 C_{bulk}(t) - C(L_0,t) \simeq L_0^{\alpha/\nu}  f_2(t L_0^{1/\nu}) 
\end{equation}
is extracted from experimental data for the specific heat  
of the three-dimensional bulk system $C_{bulk}(t)$ and of thin films 
$C(L_0,t)$, where $L_0$ is the thickness of the film. Since the specific 
heat is the derivative of the energy density with respect to the 
temperature, $f_2(x)$ is, up to a constant factor, equal to $h''(x)$.  

To compute this factor let us start from the excess reduced free energy 
density:
\begin{equation}
 f(t,L_0) - f_{bulk}(t) \simeq L_0^{-3} h(x)
\end{equation}
where $t = T/T_{\lambda}-1$ and $x=t [L_0/\xi_0]^{1/\nu}$. 
Note that here, as long as the free energy density and $L_0^{-3}$ are 
measured in the same units, $h(x)$ is uniquely defined; there is no 
ambiguous factor.

The excess energy
density is given by the derivative with respect to $\beta=1/(k_B T)$. 
Hence
\begin{equation}
 E(t,L_0) - E_{bulk}(t) =
 - L_0^{-3} [L_0/\xi_0]^{1/\nu} \frac{1}{k_B T_{\lambda} } \beta^{-2}
 \simeq - L_0^3 [L_0/\xi_0]^{1/\nu} k_B T_{\lambda}  
\end{equation}
where we have approximated $\beta^{-2} \simeq k_B^2 T_{\lambda}^2$ in 
the neighbourhood of the $\lambda$-transition. The specific heat as defined 
in the experiments is given by the derivative of the energy density with 
respect to the temperature. Hence
\begin{equation}
\label{richtig}
C_{bulk}(t) - C(t,L_0) 
  \simeq  k_B L_0^{-3} [L_0/\xi_0]^{2/\nu} h''(x)  \;.
\end{equation}

The results for the specific heat of the experiment are given in
$J \mbox{mol}^{-1} K^{-1}$. These we convert into  $k_B \AA^{-3}$  to get 
the same units on both sides of equation~(\ref{richtig}). Note that the 
thickness of the films in \cite{KiMeGa99,KiMeGa00} is quoted in $\AA^{-3}$.
To this end we need the  density
$\rho_{\lambda} = 146.1087 \; \mbox{kg/m}^3$ \cite{KeTa64} of $^4$He
at the $\lambda$-transition, the molar weight 
$4.0026\ldots \mbox{g} \; \mbox{mol}^{-1}$ of $^4$He
and the Boltzmann  constant 
$k_B=1.38065 \ldots \times 10^{-23} \mbox{J} \mbox{K}^{-1}$.
This amounts to the factor $0.00264\ldots J^{-1} \mbox{mol} K  k_B \AA^{-3}$.
In \cite{Gaspariniweb} the data are given as a function of the reduced 
temperature  $t=1-T/T_{\lambda}$, where $T_{\lambda}=2.17 \ldots K$. In 
order to plot them as a function of $x=t [L_0/\xi_0]^{1/\nu}$  we have
used $\xi_0 = 1.422(5) \AA$ which we \cite{myheat} have computed from the
amplitude of the bulk specific heat of $^4$He at vapour pressure 
\cite{lipa2003} and the universal amplitude ratio $R_{\alpha}$ 
\cite{myamplitude}.  

\begin{figure}
\begin{center}
\scalebox{0.62}
{
\includegraphics{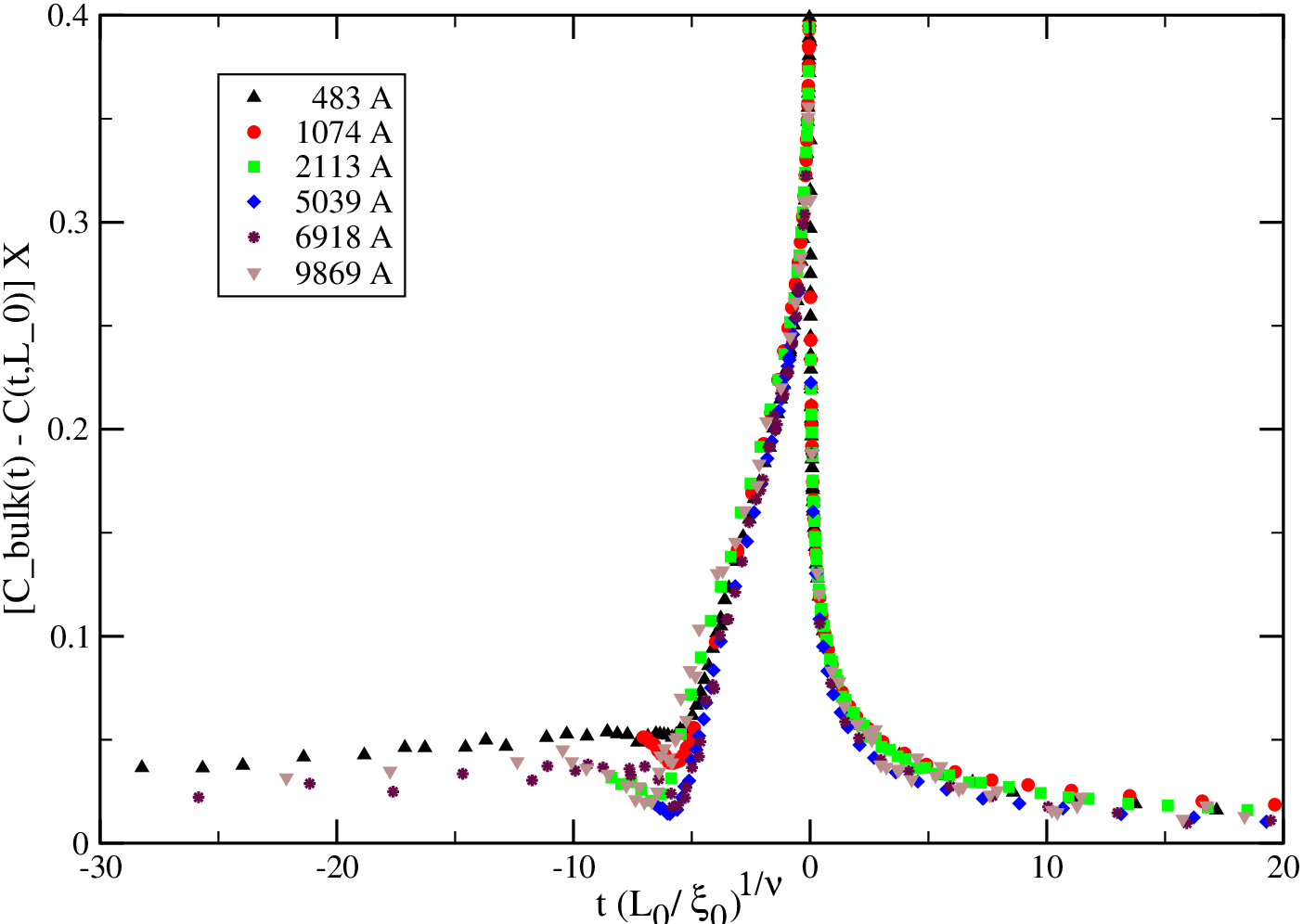}
}
\end{center}
\caption{
\label{heatplot}  We plot $[C_{bulk}(t) - C(L_0,t)] [L_0/\xi_0]^{-\alpha/\nu}$
as a function of $x=t [L_0/\xi_0]^{1/\nu}$. The data for thin films of 
$^4$He of thicknesses 483, 1074, 2113, 5039, 6918  and $9869 \AA$ obtained 
in \cite{KiMeGa99,KiMeGa00} are taken from \cite{Gaspariniweb}. Note that
at the critical point $t=0$ the finite size scaling function assumes the 
value $\approx 3.05$. For a discussion see the text.
}
\end{figure}

In figure \ref{heatplot} we have plotted
\begin{equation}
[C_{bulk}(t) - C(L_0,t)] k_B^{-1} L_0^3 [L_0/\xi_0]^{-2/\nu}  
\end{equation}
as a function of $x=t [L_0/\xi_0]^{1/\nu}$. To this end we have used the 
data given in \cite{Gaspariniweb} for the thicknesses
483, 1074, 2113, 5039, 6918  and $9869 \AA$.
\footnote{It would be interesting to repeat the analysis for other 
data sets as e.g. those of \cite{Lietal00}.}

Note that  $h''(0) \approx 3.05$ as can be obtained 
from the results of section 4.3 of \cite{myheat}. 
For $x \gtrapprox -5$ the curves obtained from different thicknesses
of the film fall reasonably well on top of each other. 
It has been noticed \cite{GaKiMoDi08} that for $x \lessapprox -5$,
in particular in the neighbourhood of the minimum, the results obtained 
for  different thicknesses differ by quite large factors.  In \cite{myheat}
we have computed the specific heat and the scaling function $f_2$ 
starting from the data for the energy density discussed above.
We find that for  $x \lessapprox -5$ our final result is clearly smaller 
than the experimental ones \cite{KiMeGa99,KiMeGa00}.

\begin{figure}
\begin{center}
\scalebox{0.62}
{
\includegraphics{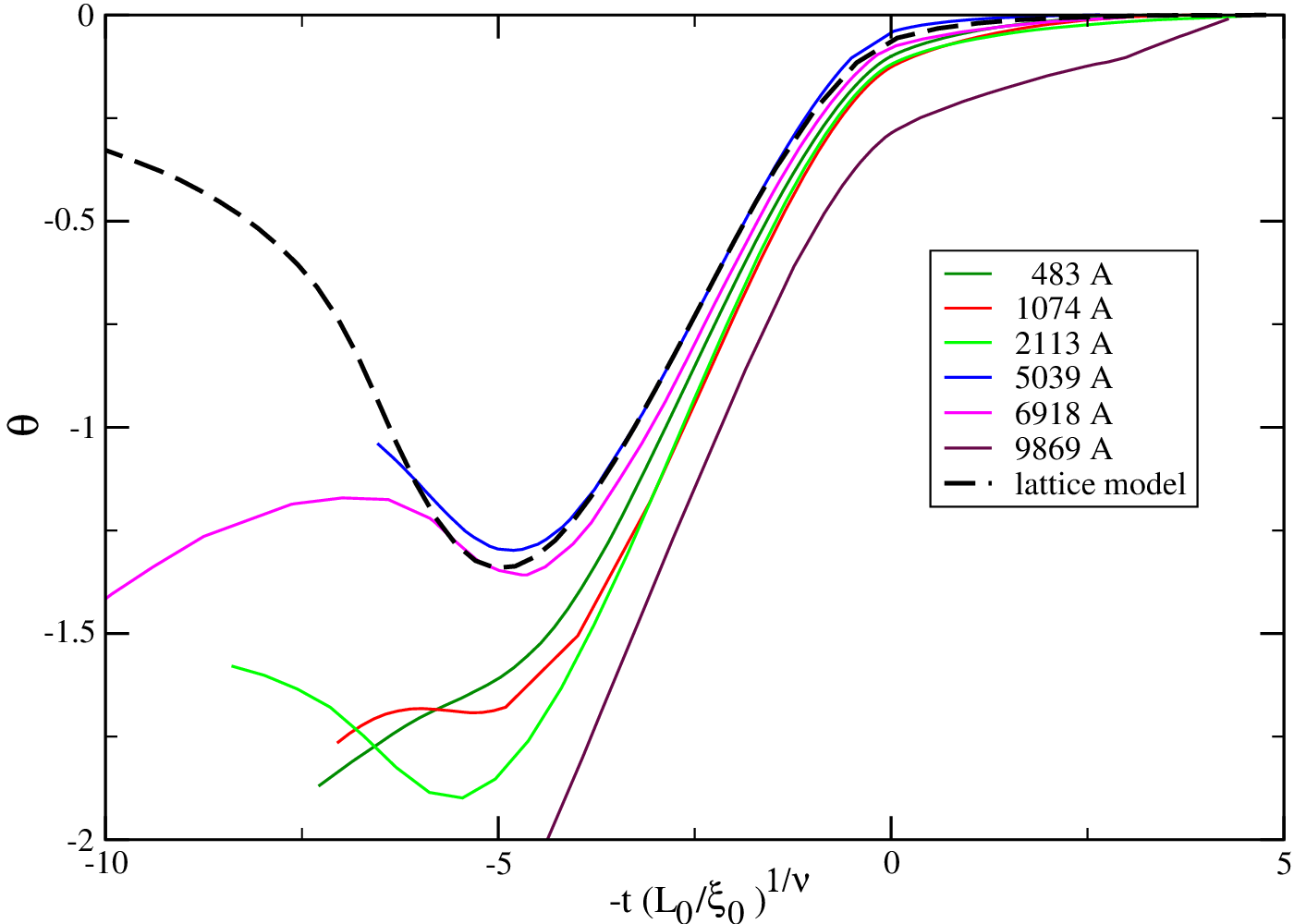}
}
\end{center}
\caption{
\label{experimentalm}  We plot results for the scaling function
$\theta$ of the thermal Casimir force. These results were obtained by 
numerically integrating experimental data \cite{KiMeGa99,KiMeGa00,Gaspariniweb}
for the excess specific heat of thin films of $^4$He near the 
$\lambda$-transition. For comparison we give the result obtained in 
\cite{myCasimir}. For a discussion see the text.
}
\end{figure}

Starting from the results  for the finite size scaling function $h''(x)$
obtained from different thicknesses we have computed the scaling function 
$h'(x)$. 
To this end, we have applied the trapezoidal rule. Similar to the 
previous section, we have started the integration at $x_0 \approx 4$
in the high temperature phase. As starting value we have taken
$h'(x_0)=\frac{x}{2 \nu-1} h''(x_0)$. Again we have integrated $h'(x)$ using 
the trapezoidal rule to get $h(x)$. Here we have taken the same value 
for $x_0$ as above and $h(x_0)=\frac{x}{2 \nu}  h'(x_0)$. 
Using these results for $h'(x)$ and 
$h(x)$  we have computed $\theta(x)$ which we have plotted in figure
\ref{experimentalm}.

In order to check the effect of errors due to the finite step-size of the 
integration, we have repeated the integration, skipping every second value
of $x$. 
In order to check the effect of the singularity of $h'(x)$  we have 
fitted the data for the specific heat in the neighbourhood of the 
transition with the ansatz 
\begin{equation}
\label{ansatz0}
 h''(x) = 3.05  + c_{\pm} |x|^{-\alpha}  
\end{equation}
Then we have integrated the ansatz in the neighbourhood of the transition
and compared it with the corresponding result from the trapezoidal rule.
We find that the numerical results only change little and the conclusions
drawn below are not effected.

Let us now discuss the results that we have obtained:
For $483 \AA$ the curve is monotonically decreasing with decreasing $x$; in 
particular no minimum of the function can be observed.
For $1074 \AA$ a shallow minimum occurs at $x  \approx - 5.3$; for 
$x   \lessapprox 6.1$ the function is decreasing again with  decreasing $x$.
For $2113 \AA$ we see a clear minimum at $x \approx - 5.5$; However 
the value of the minimum is clearly smaller than the one of \cite{myCasimir}.
In the case of $5039 \AA$ and $6918 \AA$  we find a quite good match 
with our result \cite{myCasimir} down to $x \approx - 7$. 
For $6918 \AA$ the minimum is located at $x \approx  -4.8$ and the 
value of the minimum is $\theta \approx -1.3$.
For $6918 \AA$ the minimum occurs at $x \approx 4.65$ and the value of the 
minimum is $\theta \approx -1.36$. 
For $5039 \AA$ no data for  $x<-7$ are available.
For $6918 \AA$ the curve is decreasing again for $x \lessapprox -6.7$ with 
decreasing $x$. 
Up to here the expectation that with increasing thickness 
of the film the result converges toward the universal scaling function 
is fulfilled.  However for the largest thicknesses studied, $9869 \AA$
even the worst mismatch is found.  The curve is monotonically decreasing 
with decreasing $x$ and even for $x \gtrapprox 5$ there is quite large 
mismatch with our result for $\theta$ \cite{myCasimir}. 
Playing around with the data, one finds that smaller values of $h''(x)$ 
for $x \lessapprox -5$ are needed to avoid that $\theta$ is decreasing 
with decreasing $x$ for $x \lessapprox -5$.   One should note that 
eq.~(\ref{important})
only holds for the scaling limit. Therefore the observations made here are
not an indication that the experimental data are effected by an error. 
They can also be explained by corrections to the scaling behaviour. 
This is at least true for the smaller thicknesses like $483 \AA$.  The 
result for $9869 \AA$ is more puzzling.
One should note that the analysis presented in this section 
does not depend on the type of boundary conditions that is realized in the 
experiment.  But we think that the behaviour of $\theta(x)$ for 
$x \lessapprox -5$ obtained here can not be explained by
different boundary conditions from those used in the study 
of the lattice model \cite{myheat,myCasimir}. 

\section{Summary and Conclusions}
We have studied the relation of the excess specific heat, the excess 
energy and the thermodynamic Casimir force in thin films in the 
three-dimensional XY universality class. To this end 
we have exploited the relation~(\ref{important})
\begin{equation}
 \theta(x) = 2 h(x) -\frac{x}{\nu} h'(x) 
\end{equation}
among the finite size scaling functions $\theta(x)$ of the thermodynamic 
Casimir force, $h(x)$ of the excess free energy and $h'(x)$ of the 
excess energy. We have analysed data obtained for the energy 
density of the improved two-component $\phi^4$ model on the simple cubic lattice
and experimental results for the specific heat of thin films of 
$^4$He near the $\lambda$-transition \cite{KiMeGa99,KiMeGa00}.

As a first exercise we have computed the functions $h'(x)$
and $h(x)$ for a finite lattice of the size $L^3$ with 
periodic boundary conditions in all directions.  This way we avoid 
potential problems related with the Kosterlitz-Thouless phase transition 
of the thin film and corrections caused by Dirichlet boundary conditions.
Indeed we find a good collapse of the data already for rather small lattice 
sizes $L=8$, $16$ and $32$.

Next we have repeated  the same exercise for films with free boundary 
conditions using the data for the excess energy density obtained in 
\cite{myheat}. Here we find a huge mismatch between the thicknesses 
$L_0=8$, $16$ and $32$.  Replacing $L_0$ by $L_{0,eff}=L_0+L_s$ with 
$L_s=1.02(7)$ \cite{myKTfilm} does not remove this discrepancy. 
We argue that the non-singular part of the energy density suffers from 
boundary corrections that are not  described by the same 
$L_{s}$ which accounts for the corrections in singular quantities.
Indeed, the analysis of the energy density of films up to the thickness 
$L_0=64$ at the critical temperature of the three dimensional system
results in a $L_{sns}$ for the analytic background of the energy density
that is clearly different from $L_{s}$.
Taking into account this result  we find a reasonable collapse of the 
functions obtained from $L_0=8$, $16$ and $32$ in a large range of the 
scaling variable $x$.

Computing the Casimir force from this result for $h'(x)$ we find a good
collapse for  $ -7 \lessapprox x < 4$.  
In this range of $x$ we also find a good agreement 
with the result for $\theta$ that we have obtained in \cite{myCasimir}. 
Note that in particular the minimum of the scaling function $\theta$ is 
within this range. We confirm the position $x_{min}$ and 
the value $\theta_{min}$ that we have obtained in  \cite{myCasimir}.
Next we took into account analytic corrects. The coefficients of these
corrections were computed by matching the results obtained from 
$L_0=16$ and $L_0=32$.  This way we could extend the range of 
agreement with our previous result \cite{myCasimir} down to 
$x \approx -15$. 

We have presented a viable alternative to compute the scaling function 
$\theta$ of the Casimir force.  The nice agreement with our previous 
result \cite{myCasimir} gives us further confidence in the correctness 
of the results.

Finally we have computed the scaling function $\theta$ using experimental
results for the excess specific heat \cite{KiMeGa99,KiMeGa00}.  Here 
we find a reasonable match with the theoretical results in the range 
$x \gtrapprox -5$. In particular we find evidence that the minimum 
of the scaling function $\theta$ is located at $x_{min} \approx 5$ which is
consistent with our prediction \cite{myCasimir} but slightly larger 
than $x_{min} = - 5.45(12)$ \cite{GaCh99} and
$x_{min} = - 5.7(5)$ \cite{GaScGaCh06}, where the thermodynamic Casimir 
force has been determined for films of $^4$He with thicknesses 
$\lessapprox 600 \AA$.

In the range $x \lessapprox -5$ the curves obtained from different 
thicknesses show quite different behaviour. Furthermore in this range 
$\theta(x)$ is decreasing with decreasing $x$. This  behaviour
corresponds to too large values of $h''(x)$  in the range $x \lessapprox -5$.
We think that understanding this problem requires a  detailed 
knowledge of the experiments and is therefore beyond the scope of the
present work.

\section{Acknowledgements}
This work was supported by the DFG under the grant No HA 3150/2-1.

\end{document}